\documentclass[letter,twocolumn]{jpsj3}

\usepackage{txfonts}
\usepackage{graphicx}
\usepackage{color}

\begin{document}

\title{
Surface Bound States and Spontaneous Current in Cyclic $d$-Wave Superconductors
}

\author{
Masaki \surname{Ishikawa},$^{1}$ 
\name{Yasumasa \surname{Tsutsumi},$^{2}$}
\name{Masanori \surname{Ichioka},$^{1}$} and
\name{Kazushige \surname{Machida}$^{1}$}
}
\inst{
$^{1}$Department of Physics, Okayama University, 
\address{Okayama 700-8530, Japan} \\ 
$^{2}$Condensed Matter Theory Laboratory, RIKEN, 
\address{Wako, Saitama 351-0198, Japan}
}

\date{\today}

\abst{
On the basis of Eilenberger theory, 
surface bound states and spontaneous current are studied 
in cyclic $d$-wave superconductors 
as a broken time-reversal symmetry of superconductivity  
in cubic lattice symmetry. 
We discuss how the spontaneous current and the electronic states 
depend on the orientation of the surface  
relative to the symmetry of superconductivity. 
The condition for topological Fermi arcs of zero-energy surface bound states 
to appear is identified in the complex pairing function of the cyclic $d$-wave. 
}

\kword{
surface state, 
spontaneous current, 
cyclic $d$-wave superconductivity, 
Eilenberger theory}

\maketitle


Among unconventional superconductivity, 
the broken time reversal symmetry (BTRS) of superconductivity is 
an important topic of study,  
because it has exotic properties, 
such as zero-energy surface bound states 
and spontaneous magnetic field by spontaneous current. 
In most of the previous studies of the BTRS superconductor, 
the pairing symmetry was assumed to be a chiral $p$-wave $p_x \pm{\rm i}p_y$, 
considering the cases of 
the superfluid ${\rm ^3He}$ A-phase~\cite{StoneRoy,Tsutsumi,SilaevVolovik} or 
${\rm Sr_2RuO_4}$~\cite{Luke,MackenzieMaeno}. 
The spontaneous magnetic field of a BTRS superconductor is detected by 
muon spin rotation (${\rm \mu SR}$) experiment 
in ${\rm Sr_2RuO_4}$~\cite{Luke},  
${\rm PrOs_4Sb_{12}}$~\cite{AokiPRL,AokiJPSJ}, 
${\rm LaNiC_2}$~\cite{HillierLaNiC2}, 
${\rm PrPt_4Ge_{12}}$~\cite{MaisuradzePrPt4Ge12}, 
and ${\rm LaNiGa_2}$~\cite{HillierLaNiGa2}.   
In a ${\rm \mu SR}$ experiment, ${\rm PrOs_4Sb_{12}}$, 
${\rm PrPt_4Ge_{12}}$, and ${\rm LaNiGa_2}$ show 
different types of relaxation 
curves from those of ${\rm Sr_2RuO_4}$ 
and ${\rm LaNiC_2}$. 
Therefore, there may be some variety in the types of BTRS.  
Since the crystal lattice symmetry of 
${\rm PrOs_4Sb_{12}}$ is cubic ${\rm O_h}$ 
(or more exactly, ${\rm T_h}$), the pairing symmetry may be different 
from the chiral $p$-wave of ${\rm Sr_2RuO_4}$. 
Therefore, it is important that we examine the possibility of 
a new type of pairing symmetry of BTRS other than the chiral $p$-wave, 
and study the properties of the new pairing function 
to identify the pairing symmetry. 
  
In this study, we consider the BTRS of superconductivity 
in cubic lattice symmetry. 
From the classification table of possible pairing symmetries  
under spin-orbit coupling 
on the basis of point group theory,~\cite{VolovikGorkov,SigristUeda,Ozaki}  
as a BTRS state keeping ${\rm O_h}$ symmetric superconductivity, 
we find 
$k_x^2+\omega_\pm k_y^2 +\omega_\pm^2 k_z^2$ 
in spin-singlet pairing and 
$k_x \hat{x} +\omega_\pm k_y \hat{y} +\omega_\pm^2 k_z \hat{z}$ 
in spin-triplet pairing with 
$\omega_\pm={\rm e}^{\pm{\rm i}2\pi/3}$. 
The former is cyclic $d$-wave pairing, 
and it is stable in weak coupling theory. 
The latter is non-unitary spin-triplet pairing, 
and it is not stable in weak coupling theory. 
Thus, the case of cyclic $d$-wave pairing is studied here. 
The possibility of cyclic $d$-wave superconductivity and 
its exotic properties have been discussed in the study of 
Bose-Einstein condensation~\cite{Semenoff,Nakeka,Kobayashi} 
and fermionic superfluidity~\cite{Adachi} in cold atomic gases.
There, the non-Abelian $\frac{1}{3}$-fractional vortex is also possible 
in an isotropic atomic gas system. 
Thus, the BTRS of cyclic $d$-wave pairing is 
one of the interesting pairing symmetries  
for the theoretical study of topological superconductivity.

Studies of the surface state are important to identify the pairing symmetry 
of unconventional superconductivity. 
For example, in a $d_{x^2-y^2}$-wave superconductor, 
the surface bound state depends on the surface orientation, 
and zero-energy surface bound states appear 
for the (1,1,0) surface.~\cite{Tanaka,KashiwayaTanaka} 
The characteristic dispersion relation of the surface bound state 
reflects the pairing symmetry of the bulk superconductivity. 
In the $d_{xy}$-wave pairing and the $p_x$-wave pairing, 
we have flat dispersion of the surface bound states 
for the (1,0,0) surface.~\cite{KashiwayaTanaka,Sato}  
In superfluid ${\rm ^3He}$, surface bound states show 
topological Fermi arcs for the A-phase  
and a Majorana cone for the B phase.~\cite{SilaevVolovik,Tsutsumi} 
Therefore, studies of the surface bound state are also necessary for 
the complex pairing function of cyclic $d$-wave superconductivity.

The purpose of this study is to clarify the properties of 
surface bound states in a cyclic $d$-wave superconductor, 
as another example of the BTRS state, on the basis of Eilenberger theory. 
Since the structure of surface bound states depends on the relative angle 
of the surface and the pairing symmetry, 
we study the surface-orientation dependence of spontaneous current 
and electronic states in the surface bound state. 
We also discuss the condition under which the zero-energy surface bound state 
appears in the cyclic $d$-wave superconductivity. 


In our study, wave numbers are denoted as 
$(k_a,k_b,k_c)$ for crystal coordinates. 
Under an ${\rm O_h}$ crystal field, we assume the pairing function  
to be the cyclic $d$-wave given by 
\begin{eqnarray}&& 
\phi_\pm({\bf k}) 
=\sqrt{\frac{5}{2}}(k_c^2 + \omega_\pm k_a^2 +\omega_\pm^2 k_b^2)
\nonumber \\ && 
=\frac{1}{\sqrt{2}} \left(\frac{\sqrt{5}}{2}(3k_c^2-1)
\pm {\rm i}\frac{\sqrt{15}}{2}(k_a^2-k_b^2) \right).   
\label{eq:OP}
\end{eqnarray} 
Therefore, $\phi_-({\bf k})=\phi_+^\ast({\bf k})$. 
In Eq. (\ref{eq:OP}), ${\bf k}$ is mapped on the Fermi sphere  
and normalized as $k_a^2+k_b^2+k_c^2=1$.  
From Eq. (\ref{eq:OP}), the cyclic $d$-wave is a combination of 
$d_{x^2-y^2}$-wave and $d_{3z^2-1}$-wave components.  
The amplitude $|\phi_\pm({\bf k})|$ has ${\rm O_h}$ symmetry 
with 8 point nodes in (1,1,1) and equivalent directions. 
The phase of $\phi_\pm({\bf k})$ has ${\rm T_h}$ symmetry. 
For $120^\circ$ rotation around the (1,1,1)-axis 
($k_a \rightarrow k_b$, $k_b \rightarrow k_c$, $k_c \rightarrow k_a$), 
$\phi_+ \rightarrow \omega^2 \phi_+$. 
For $90^\circ$ rotation around the (0,0,1)-axis 
($k_a \rightarrow k_b$, $k_b \rightarrow -k_a$), 
$\phi_+ \rightarrow \phi_-(\ne \phi_+)$. 
In the calculation of the surface state, we use the coordinates  
$(k_x,k_y,k_z)$ and $(x,y,z)$,  
where the $x$-axis is fixed to be perpendicular to the surface at $x=0$. 
$(k_x,k_y,k_z)$ is obtained by rotational transformation from $(k_a,k_b,k_c)$.
The $x$-direction is defined as 
$(\cos\theta \cos\phi, \cos\theta \sin\phi, \sin\theta)$ 
in the crystal coordinates. 

Surface states are calculated on the basis of Eilenberger theory, 
following the method used to study the surface states of 
superfluid ${\rm ^3He}$~\cite{Tsutsumi}. 
Since we consider the case of spin-singlet pairing here, 
the transport-like Eilenberger equation 
is reduced to the $2 \times 2$ matrix form 
\begin{eqnarray} && 
-{\rm i}{\bf v}\cdot\nabla \hat{g}
=
\frac{1}{2}
\left[ \left(
 \begin{array}{cc} {\rm i}\omega_n & -\Delta({\bf k},{\bf r}) \\ 
                   \Delta^\ast({\bf k},{\bf r}) & -{\rm i}\omega_n \\ 
 \end{array} \right), 
 \hat{g} \right] ,
\label{eq:Eilenberger}
 \end{eqnarray}
for quasi-classical Green's functions 
\begin{eqnarray}
 \hat{g}
=-{\rm i}\pi \left( \begin{array}{cc} 
  g({\bf k},{\bf r},\omega_n) & {\rm i}f({\bf k},{\bf r},\omega_n) \\ 
  -{\rm i}f^\dagger({\bf k},{\bf r},\omega_n) & -g({\bf k},{\bf r},\omega_n) \\ 
  \end{array}  \right)
\end{eqnarray} 
with the normalization condition $\hat{g}^2=-\pi^2 \hat{1}$. 
$\omega_n$ is the Matsubara frequency. 
We consider the case where an external magnetic field is not applied. 
For simplicity, we neglect the contribution of vector potentials. 
The pair potential takes the general form 
\begin{eqnarray} 
\Delta({\bf k},{\bf r})
=\Delta_+({\bf r}) \phi_+({\bf k}) 
+\Delta_-({\bf r}) \phi_-({\bf k}) . 
\end{eqnarray} 
We treat $\Delta_+$ as the dominant component and 
$\Delta_-$ as the induced component near the surface, 
so that $\Delta_-=0$ and $\Delta_+$ is a bulk value 
when $x \rightarrow \infty$. 
Since we assume a spherical Fermi surface, 
the normalized Fermi velocity is ${\bf v}={\bf k}$. 
We assume that the surface condition is specular, and 
a quasi-particle with $(-k_x,k_y,k_z)$ is reflected to $(k_x,k_y,k_z)$ 
at $x=0$.     
We solve the Riccati equation derived from the Eilenberger equation 
[Eq. (\ref{eq:Eilenberger})], and 
obtain the quasi-classical Green's functions. 
Energy, temperature, and length are in units of $\pi k_{\rm B} T_{\rm c}$, 
$T_{\rm c}$, 
and $\xi_0=\hbar v_{{\rm F}}/2\pi k_{\rm B} T_{\rm c}$,  respectively. 
The pair potential is calculated by the gap equation 
\begin{eqnarray}
\Delta_\pm({\bf r})
= \lambda_0\, 2T \sum_{\omega_n >0}^{\omega_{\rm c}}
 \left\langle \phi_\pm^\ast({\bf k})  f  \right\rangle_{\bf k} , 
\label{eq:scD}
\end{eqnarray}
where 
$\langle\cdots\rangle_{\bf k}$ indicates the Fermi surface average.  
$\lambda_0=N_0g_0$ is the dimensionless pairing interaction 
defined by the cutoff energy $\omega_{\rm c}$ as 
$1/\lambda_0 = \ln T+2\,T\sum_{\omega_n>0}^{\omega_{\rm c}}\,\omega_n^{-1}$.
We carry out calculations using the cutoff $\omega_{\rm c}=40 k_{\rm B}T_{\rm c}$. 
The calculations of Eqs. (\ref{eq:Eilenberger}) and (\ref{eq:scD}) 
are iterated at $T=0.2T_{\rm c}$ until self-consistent results are obtained. 

The current density of spontaneous current is given as  
\begin{eqnarray}
{\bf J}({\bf r})  
=j_0  2T 
\sum_{\omega_n > 0}\left\langle {\bf v} \,{\rm Im}\,g\right\rangle_{\bf k} , 
\label{eq:scH}
\end{eqnarray}
and produces the spontaneous magnetic field.  
$j_0=8 \pi {\rm e} N_0 /c$, and 
$N_0$ is the density of states at the Fermi level. 
The total current of the surface bound states is given by the integral 
\begin{eqnarray}
{\bf J}_{\rm total}=\int_0^\infty {\bf J}({\bf r}){\rm d}x .
\end {eqnarray}
The local density of states (LDOS) is obtained as 
\begin{eqnarray}
N(E,{\bf r})=\langle N(E,{\bf k},{\bf r}) \rangle_{\bf k}
=N_0 \langle 
{\rm Re}g({\bf k},{\bf r},{\rm i}\omega_n \rightarrow E+{\rm i}\eta) 
\rangle_{\bf k} , 
\end{eqnarray}
where we use the solution of the Eilenberger equation 
[Eq. (\ref{eq:Eilenberger})] for 
the real energy $E$ in the self-consistently obtained pair potential. 
We use a small smearing of $\eta=0.007$. 
$N(E,{\bf k},{\bf r})$ is the ${\bf k}$-resolved LDOS. 
 

\begin{figure}[t]
\begin{center}
\includegraphics[width=8.5cm]{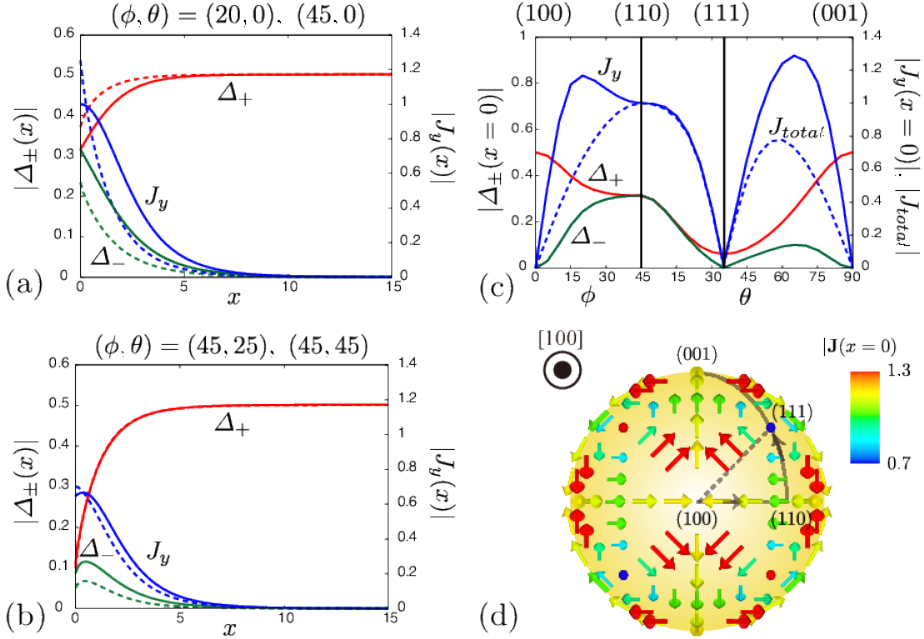}

\vspace{-0.5cm}
\end{center}
\caption{\label{fig1}
(Color online) 
(a) Depth dependence of pair potential $|\Delta_+(x)|$, $|\Delta_-(x)|$,  
and spontaneous current $|J(x)|$ as a function of distance $x$ from 
the surface. 
Solid lines are for the surface orientation $(\phi,\theta)=(45^\circ,0)$  
and dashed lines are for $(\phi,\theta)=(20^\circ,0)$.
$|\Delta_\pm|$ is in the unit of $\pi k_{\rm B} T_{\rm c}$.
$|{\bf J}(x)|$ and $|{\bf J}_{\rm total}|$ are, respectively, normalized by 
the values $|{\bf J}(x=0)|$ and $|{\bf J}_{\rm total}|$ 
for $(\phi,\theta)=(45^\circ,0)$. 
(b) Same as (a), but $(\phi,\theta)=(45^\circ,25^\circ)$ for solid lines  
and $(\phi,\theta)=(45^\circ,45^\circ)$ for dashed lines.  
(c) Surface orientation $(\phi,\theta)$ dependence of the surface state 
$|\Delta_+(x=0)|$, $|\Delta_-(x=0)|$, $|{\bf J}(x=0)|$ (solid lines), and 
$|{\bf J}_{\rm total}|$ (dashed lines). 
The horizontal axis indicates that the surface orientation 
changes from (1,0,0) to (1,1,0) [$\theta=0$ and $\phi=0 \rightarrow 45^\circ$] 
and from (1,1,0) to (0,0,1) [$\phi=45^\circ$ and 
$\theta=0 \rightarrow 90^\circ$]. 
(d) Schematic plot of spontaneous current 
at the surface of a large spherical sample. 
The line indicates the path of the horizontal axis in (c). 
The path (1,1,1)-(0,0,1) is equivalent to the path (1,1,1)-(1,0,0). 
}
\end{figure}
\begin{figure}[tb]
\begin{center}
\includegraphics[width=5.6cm]{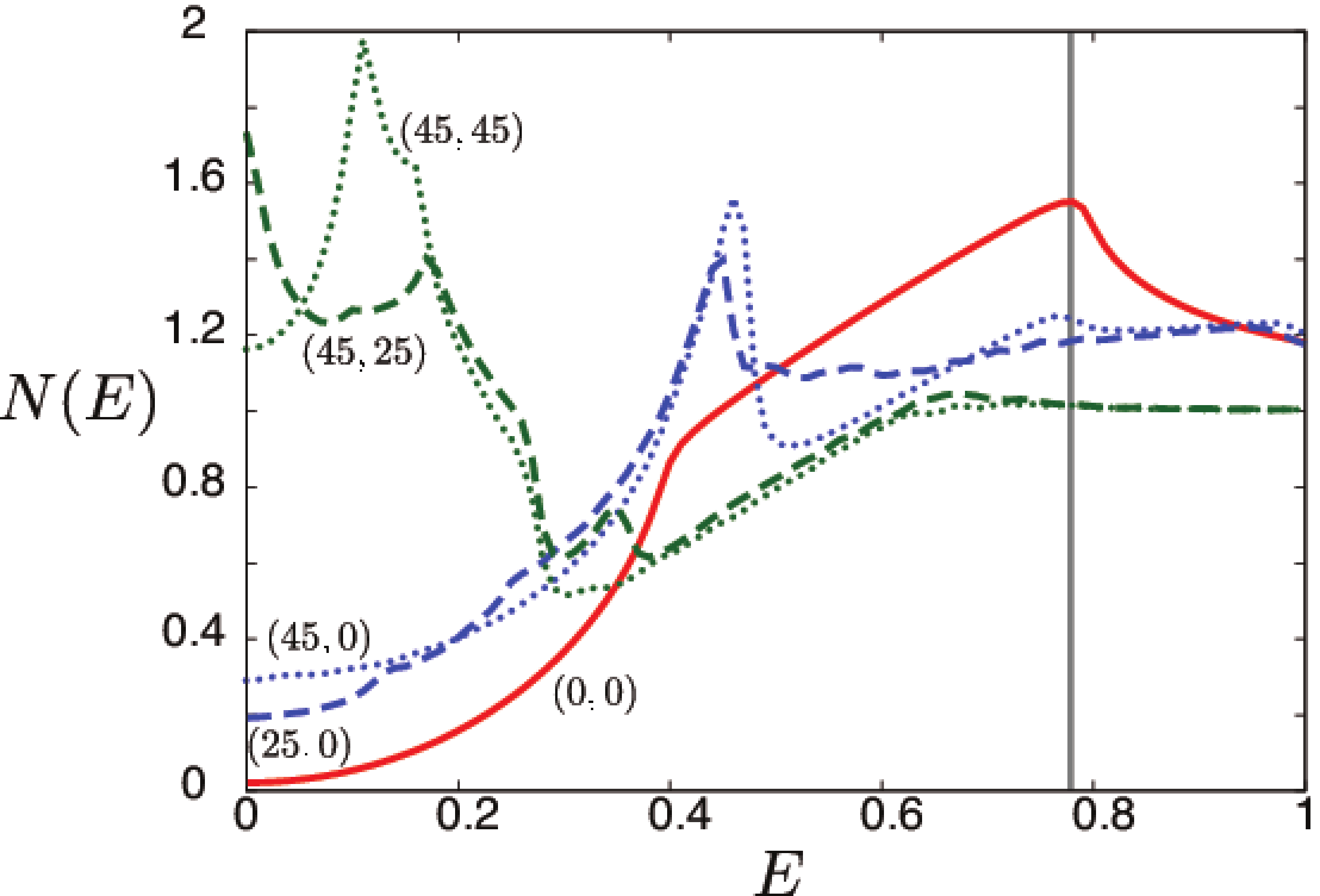}

\vspace{-0.5cm}
\end{center}
\caption{\label{fig2}
(Color online) 
LDOS $N(E,{\bf r})$ at surface $x=0$ with various surface orientations,  
$(\phi,\theta)=(0,0)$, $(25^\circ,0)$, $(45^\circ,0)$,  
$(45^\circ,25^\circ)$, and $(45^\circ,45^\circ)$. 
}
\end{figure}
\begin{figure*}[tb]
\includegraphics[width=17.0cm]{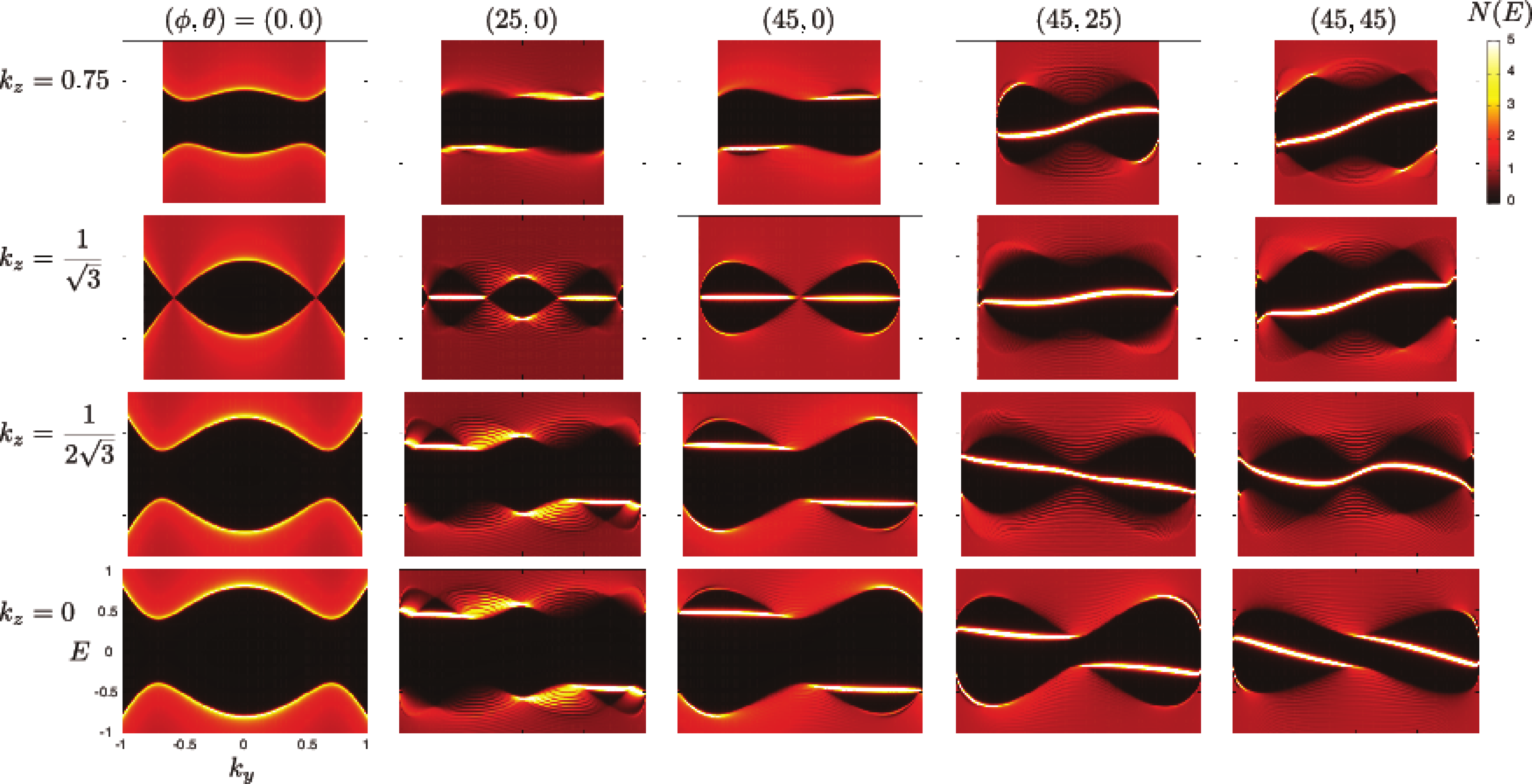}

\vspace{-8.8cm}
\hspace{1.2cm} (a) 
\hspace{2.5cm} (b) 
\hspace{2.5cm} (c) 
\hspace{2.5cm} (d) 
\hspace{2.5cm} (e)
\vspace{8.0cm}

\caption{\label{fig3}
(Color online) 
${\bf k}$-resolved LDOS $N(E,{\bf k},{\bf r})$ at surface $x=0$ 
as functions of $k_y$ and $E$ 
for $k_z=0$, $1/(2\sqrt{3})$, $1/\sqrt{3}$, and $0.75$. 
$k_x=(1-k_y^2-k_z^2)^{1/2}$. 
Various surface orientations,   
$(\phi,\theta)=(0,0)$ (a), $(25^\circ,0)$ (b), $(45^\circ,0)$ (c),  
$(45^\circ,25^\circ)$ (d), and $(45^\circ,45^\circ)$ (e), are presented. 
}
\end{figure*}

First, in Figs. \ref{fig1}(a) and \ref{fig1}(b), 
we present the self-consistent solution of 
the pair potential and spontaneous current 
for some surface orientations. 
At the surface region, when the length is of the order of the coherence length, 
$\Delta_+$ is suppressed, and the induced $\Delta_-$ accompanies 
the $y$-component of the spontaneous current. 
These behaviors depend on the surface orientation. 
For the (1,1,0) surface with $(\phi,\theta)=(45^\circ,0)$, 
since the $d_{x^2-y^2}$-wave component in the pairing function changes sign 
in the quasi-particle reflection at the surface,  
$|\Delta_+|=|\Delta_-|$ at the surface so that 
the $d_{x^2-y^2}$-wave component vanishes there. 
On the other hand, 
for the (1,0,0) surface with $(\phi,\theta)=(0,0)$, 
$\Delta_-=0$ and ${\bf J}=0$ even near the surface, 
and the surface state is the same as the bulk state. 

To confirm how the surface state depends on the surface orientation, 
in Fig. \ref{fig1}(c),  
we plot the surface values $|\Delta_\pm(x=0)|$, $|{\bf J}(x=0)|$,  
and $|{\bf J}_{\rm total}|$ 
along the trajectory (1,0,0)-(1,1,0)-(1,1,1)-(0,0,1). 
Along (1,0,0)-(1,1,0), 
$|\Delta_+(x=0)|$ decreases and $|\Delta_-(x=0)|$ increases.  
$|{\bf J}(x=0)|$ increases and is maximum at $\phi=20^\circ$. 
However, the total current $|{\bf J}_{\rm total}|$ monotonically 
increases until $\phi=45^\circ$. 
This is because the surface region in which spontaneous current appears  
becomes wider as $\phi$ approaches $45^\circ$, 
as seen in Fig. \ref{fig1}(a). 
Along (1,1,0)-(1,1,1), $|\Delta_\pm(x=0)|$ and current decrease.  
For the (1,1,1) surface of the point node direction, 
$\Delta_-(x=0)=0$ and $J(x)$ vanishes.    
Along (1,1,1)-(0,0,1), which is equivalent to (1,1,1)-(1,0,0), 
$|\Delta_+(x=0)|$ increases monotonically. 
$|\Delta_-(x=0)|$ and current decrease after they increase.  

In order to understand the flow of spontaneous current ${\bf J}$, 
we consider the surface current in a large spherical superconductor 
of cyclic $d$-wave pairing. 
As in Fig. \ref{fig1}(d), 
at the point node direction (1,1,1) and the equivalent points, ${\bf J}=0$. 
Around the node points (1,1,1), $(1,-1,-1)$, $(-1,1,-1)$, and $(-1,-1,1)$, 
${\bf J}$ flows clockwise.
Around other node points $(-1,-1,-1)$, $(-1,1,1)$, $(1,-1,1)$, and $(1,1,-1)$,  
${\bf J}$ flows counterclockwise. 
The direction of the flow is related to 
the angular momentum of the Cooper pairs 
via the phase winding of the spherical harmonic function, $Y_{2,m}$,  
on the Fermi sphere. 
The pairing function is expressed as 
$\phi_+ \propto Y_{2,-2} -{\rm i}\sqrt{2}Y_{2,1}$ when $z \parallel (1,1,1)$  
and as 
$\phi_+ \propto Y_{2,2} +\sqrt{2}Y_{2,-1}$ when $z \parallel (1,-1,1)
$.~\cite{Adachi}  
The spontaneous current at the surface is induced 
by the change in the angular momentum of Cooper pairs when they are 
reflected at the surface. 
By summing the spontaneous currents 
around neighboring point nodes, 
${\bf J}$ is enhanced at (1,1,0) and the equivalent positions. 
On the other hand, at (1,0,0) and the equivalent positions, 
${\bf J}$ is canceled to zero 
by summing neighboring spontaneous currents. 
These current distributions may be reflected in the spontaneous 
magnetic field distribution that is expected to be observed.   



Next, we discuss electronic states at the surface. 
In Fig. \ref{fig2}, 
we present the LDOS $N(E,x=0)$ for some cases of surface orientation. 
The LDOS is expected to be observed by tunneling spectroscopy at the surface. 
For the (1,0,0) surface with $(\phi,\theta)=(0,0)$, 
surface states have the same electronic states as those in the bulk. 
There, $N(E)\propto E^2$ at low $E$ owing to point node excitations. 
The gap edge at $E \sim 0.8$ comes from the maximum of 
$|\Delta_+ \phi_+({\bf k})|$ at $(k_a,k_b,k_c) \propto (1,0,0)$. 
The hump at $E \sim 0.4$ corresponds to 
the saddle points of 
$|\Delta_+ \phi_+({\bf k})|$ at $(k_a,k_b,k_c) \propto (1,1,0)$.  

With increasing $\phi$ from (1,0,0) to (1,1,0), 
low-energy surface bound states appear, including zero-energy states. 
The peak at $E \sim 0.8$ decays, and a new peak appears at $E \sim 0.5$. 
In the cases of surface orientations 
$(\phi,\theta)=(45^\circ, 25^\circ)$ and $(45^\circ, 45^\circ)$, 
the LDOS $N(E,x=0)$ has high intensity at low energy, 
because both $\Delta_+(x=0)$ and $\Delta_-(x=0)$ are 
largely suppressed at the surface. 


In order to understand the structures of the surface bound state and 
spontaneous current, we study the surface orientation dependence of 
the ${\bf k}$-resolved LDOS $N(E,{\bf k},{\bf r})$ at the surface $x=0$, 
as presented in Fig. \ref{fig3}, 
to examine the dispersion relation of surface bound states. 
For the (1,0,0) surface with $(\phi,\theta)=(0,0)$, 
the surface state is the same as the bulk state. 
Thus, $N(E,{\bf k},x=0)$ shows the gap structure of the pairing function 
$|\Delta_+ \phi_+({\bf k})|$. 
In the panel for $k_z=0$, we see the maximum of the gap 
$|\Delta_+ \phi_+({\bf k})|$ 
to be $E \sim 0.8$ at $k_y=0$, $\pm 1$, 
and the saddle point energy $E \sim 0.4$ at $k_y =\pm 1/\sqrt{2}$. 
In the panel for $k_z=1/\sqrt{3}$, zero energy states appear  
due to the point nodes at $k_y=\pm 1/\sqrt{3}$. 

With increasing $\phi$ from (1,0,0) to (1,1,0), 
in-gap states of the surface bound states appear. 
In the panel for $k_z=1/\sqrt{3}$, 
flat dispersions of zero-energy modes exist at $0.28 < |k_y| < 0.77$ 
for the surface with $(\phi,\theta)=(25^\circ,0)$. 
For the (1,1,0)-surface with $(\phi,\theta)=(45^\circ,0)$, 
zero-energy modes extend to all $k_y$ at $k_z=1/\sqrt{3}$. 
This is a property of the $d_{x^2-y^2}$-wave component, 
because the $d_{3z^2-1}$-wave component vanishes at $k_z=1/\sqrt{3}$. 
In Figs. \ref{fig3}(b) and \ref{fig3}(c), 
the flat dispersions are raised to finite energies at $k_z \ne 1/\sqrt{3}$. 
When $k_z > 1/\sqrt{3}$ ($k_z < 1/\sqrt{3}$), 
they appear at $E>0$ ($E<0$) for $k_y>0$ and 
at $E<0$ ($E>0$) for $k_y<0$. 
Since a negative $E$ indicates an occupied state, 
the imbalance of negative $E$ states of $k_y>0$ and $k_y<0$ 
induces the spontaneous current. 
Since the contribution of the negative $E$ state of $k_y>0$ at $k_z<1/\sqrt{3}$ 
is stronger than that of $k_y<0$ at $k_z>1/\sqrt{3}$, 
this difference causes the spontaneous current 
to flow in the positive $y$ direction. 
 
When the surface orientation changes from (1,1,0) to (0,0,1) upon  
increasing $\theta$ at $\phi=45^\circ$, 
the flat dispersions in Fig. \ref{fig3}(c) become dispersive, 
as shown in Figs. \ref{fig3}(d) and \ref{fig3}(e). 
We see the reconnection of the dispersion curve between $\theta=0$ 
and $25^\circ$ in panels for 
$k_z=1/(2\sqrt{3})$ and $0.75$. 
The reconnection also occurs between $\theta=25^\circ$ and $45^\circ$ 
in the panel for $k_z=0$. 
Zero-energy states appear at $k_y=0$ in the upper three panels in both 
Figs. \ref{fig3}(d) and \ref{fig3}(e), 
indicating flat dispersion on part of the line $k_y=0$.  
In Fig. \ref{fig3}(e), we also see zero-energy states at $k_y \ne 0$ 
in the panels for $k_z=0$ and $k_z=1/(2\sqrt{3})$. 

\begin{figure}[bt]
\begin{center}
\includegraphics[width=8.5cm]{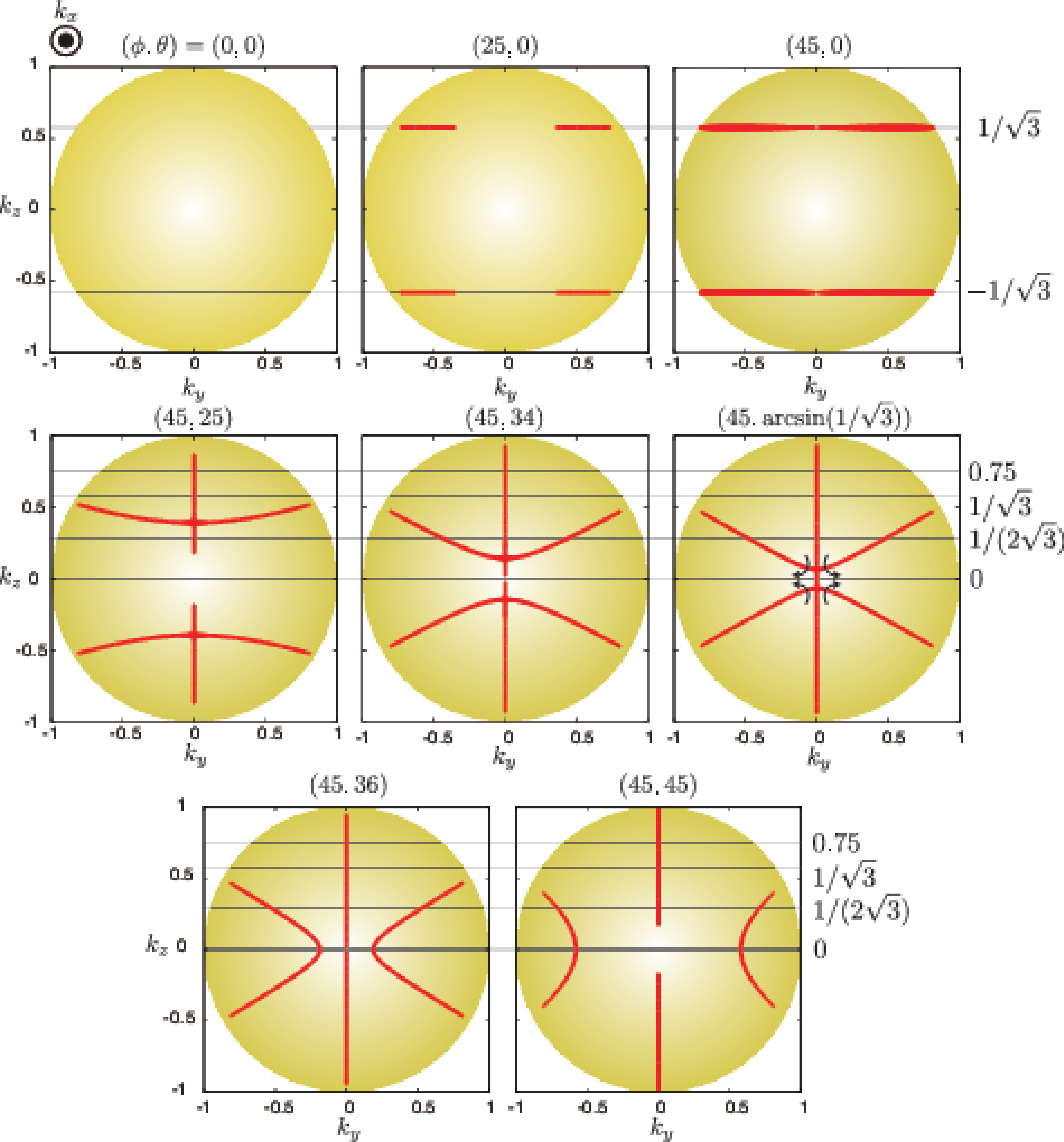}

\vspace{-0.5cm}
\end{center}
\caption{\label{fig4}
(Color online) 
Bold lines indicate wave numbers $(k_y,k_z)$ 
of topological Fermi arcs
satisfying the condition for zero-energy states, Eq. (\ref{eq:E0}).
$k_x=(1-k_y^2-k_z^2)^{1/2}$. 
Various surface orientation 
cases, $(\phi,\theta)=(0,0)$, $(25^\circ,0)$, $(45^\circ,0)$,  
$(45^\circ,25^\circ)$, $(45^\circ,34^\circ)$, 
$(45^\circ,{\rm arcsin}(1/\sqrt{3}))$, $(45^\circ,36^\circ)$, 
and $(45^\circ,45^\circ)$, are presented. 
The thin lines $k_z=0$, $1/(2\sqrt{3})$, 
$1/\sqrt{3}$, and 0.75 correspond to 
the horizontal axis in the panels in Fig. \ref{fig3}. 
The intersections of bold and thin lines indicate that 
the zero-energy states appear at $(k_y,k_z)$ in Fig. \ref{fig3}. 
}
\end{figure}

Lastly, we discuss the condition of ${\bf k}$ for 
zero-energy surface bound states to appear 
in the complex pairing function of cyclic $d$-wave superconductivity. 
We define the phase $\varphi_{\rm in}$ of the pairing function 
before reflection at the surface as 
$\phi_+(-k_x,k_y,k_z)=|\phi_+(-k_x,k_y,k_z)|{\rm e}^{{\rm i}\varphi_{\rm in}}$  
and the phase $\varphi_{\rm ref}$ after reflection as 
$\phi_+(k_x,k_y,k_z)=|\phi_+(k_x,k_y,k_z)|{\rm e}^{{\rm i}\varphi_{\rm ref}}$. 
As the condition under which the zero-energy surface bound states 
in Fig. \ref{fig3} are well explained, we find the relation
\begin{eqnarray}
{\rm e}^{{\rm i}\varphi_{\rm in}}=- {\rm e}^{{\rm i}\varphi_{\rm ref}}, 
\label{eq:E0}
\end{eqnarray}
i.e., 
the $\pi$-phase shift    
$\varphi_{\rm ref}-\varphi_{\rm in}=\pi$ (mod $2\pi$).~\cite{KashiwayaTanaka}  
Equation (\ref{eq:E0}) can be applied even 
when $|\phi_+(-k_x,k_y,k_z)| \ne |\phi_+(k_x,k_y,k_z)|$.  

Wave numbers ${\bf k}$ satisfying Eq. (\ref{eq:E0}) 
are presented in Fig. \ref{fig4}. 
These topological Fermi arcs 
are terminated at the point node directions.~\cite{SilaevVolovik}  
For the (1,0,0)-surface with $(\phi,\theta)=(0,0)$, 
there are no zero-energy surface states. 
For the (1,1,0)-surface with $(\phi,\theta)=(45^\circ,0)$, 
all $k_y$ values satisfy Eq. (\ref{eq:E0}) on the lines $k_y=\pm 1/\sqrt{3}$. 
When changing $\phi$ from (1,0,0) to (1,1,0), 
the region of $k_y$ in which zero-energy surface bound states appear increases. 
These reproduce the behavior of the zero-energy flat dispersion 
in the panels for $k_z=1/\sqrt{3}$ in Figs. \ref{fig3}(a)-\ref{fig3}(c). 
When changing $\theta$ from (1,1,0) to (0,0,1), 
the lines of zero energy at $k_z=1/\sqrt{3}$ are shifted to smaller $|k_z|$ 
and become diagonal lines. 
Zero-energy states also appear on the vertical line $k_y=0$. 
The endpoints of the zero-energy states 
on the vertical line are the point node directions. 
When $\theta$ approaches ${\rm arcsin}(1/\sqrt{3}) \sim 35^\circ$ 
of the (1,1,1)-surface direction, all vertical and diagonal lines 
pass through the center $(k_y,k_z)=(0,0)$. 
The reconnection of these lines occurs at this point-node wave number, 
as shown in Fig. \ref{fig4}.  
From the panel for $(\phi,\theta)=(45^\circ,25^\circ)$, 
we find that zero-energy states appear at $k_y=0$ in the plots along 
the lines $k_z=1/(2\sqrt{3})$, $1/\sqrt{3}$, and $0.75$. 
From the panel for $(\phi,\theta)=(45^\circ,45^\circ)$, 
zero-energy states are seen to also appear at $k_y \ne 0$ 
in the plots along $k_z=0$ and $1/(2\sqrt{3})$. 
These well explain the wave numbers of the zero-energy states 
in Figs. \ref{fig3}(d) and \ref{fig3}(e). 


In summary, 
we studied the surface orientation dependence of the surface states 
in the cyclic $d$-wave superconductor, 
as an example of the BTRS state in 
cubic lattice symmetric superconductivity. 
There, spontaneous currents flow around each point node direction  
if we prepare spherical samples.
We also identified the condition under which  
topological Fermi arcs of 
zero-energy surface bound states 
appear in cyclic $d$-wave superconductors. 
These are useful results for studying new types of BTRS superconductors 
other than those with chiral $p$-wave pairing. 


We thank T. Mizushima and T. Kawakami for fruitful discussions. 
This work was supported by KAKENHI 
Grants No. 21340103 and No. 24840048. 



\end{document}